\def\ii#1\ff{\textul{#1}}	

\def\bbsty#1#2#3{{\bf #1}, (#3) #2}	

\documentclass[psfig,amsmath,amssymb]{cimento_modif}
\usepackage{graphicx}\usepackage{epsfig}
\usepackage{amsmath}\usepackage{amssymb}\usepackage{amsfonts}
\usepackage{bm}
\usepackage{exscale}
\usepackage{dcolumn}    
\usepackage{color}
\usepackage{soul}
\usepackage{ulem}
\title{\scalebox{1.25}[1.6]{How nuclear jets form and disintegrate into clusters} \\[.8ex]
	   \scalebox{1.25}[1.6]{in heavy-ion collisions}}
\shorttitle{How nuclear jets form and disintegrate into clusters...}
\author{
	P.~Napolitani\from{ins:x}	\atque
	M.~Colonna\from{ins:y}	}
\instlist{
	\inst{ins:x} IPN, CNRS/IN2P3, Universit\'e Paris-Sud, Universit\'e Paris-Saclay, 91406 Orsay, France
	\inst{ins:y} INFN-LNS, Laboratori Nazionali del Sud, 95123 Catania, Italy}

\begin{document}

\maketitle

\begin{abstract}
    The most extreme deformations that can be explored in heavy-ion collisions at Fermi-energies are collimated flows of nuclear matter which recall jet dynamics.
    From microphysics to the cosmological scale, jets are rather common topologies.
    In nuclear physics, pioneering works focused on the breakup of these structures, resulting into early nuclear-fission models in analogy to the droplet formation in viscous liquids; such view became emblematic to explain surface-energy effects and surface instability by analogy with the Rayleigh instability.
    Through a dynamical approach based on the Boltzmann-Langevin equation, well adapted to out-of-equilibrium conditions, we explored the possibility that nuclear jets could arise in heavy-ion collisions from different conditions than those leading to fission or neck fragmentation, and that they can breakup from mechanisms that are almost unrelated to cohesive properties.
\end{abstract}

%
%
%
%
%
\section{Introduction: nuclear jets}

Jets are collimated streams of matter which, under specific circumstances, can disintegrate and rearrange into packets, droplets or clusters.
	Widely encountered in nature at all physical scales, jets exhibit a large variety of non-linear behaviours and rupture mechanisms from different sources of instability, depending on the type of fluid.
	In this brief report we focus on Fermi fluids~\cite{Pines1966} and some peculiar dynamical properties which can be investigated via heavy-ion collisions (HIC)~\cite{EPJAtopicalWCI2006}.
	Already ordinary fluids may manifest phenomena which are paradoxical enough to stimulate curiosity~\cite{Eggers2008}.
	For example, while water falling from a pipette gives rise to the formation of ligaments and droplets from the effect of surface tension, an identical behaviour is displayed in streams of granular material, like dry sand, in total absence of surface tension.
	Accurate descriptions of liquid-jet behaviour date back to the works of Rayleigh~\cite{Rayleigh1882} and they have been regarded as useful analogies for the first models of nuclear reactions~\cite{Griffin1976} like, for instance, fission~\cite{Vandenbosch1973} and the phenomenology of nuclear necks~\cite{Baran2005,Lionti2005,DiToro2006,Baran2012}.
	On the contrary, cluster formation in granular jets has been understood only recently as relying on correlations beyond one-body dynamics~\cite{Royer2008,Royer2009}.
	Are similar mechanisms also present in nuclear physics?

	We argue that heavy-ion collisions in the Fermi energy domain can produce extremely deformed configurations which suddenly turn into nuclear clusters~\cite{Napolitani_submitted}.
	Typically, they arise when a target nucleus is hit by a lighter projectile nucleus: while the former then turns into a heated compound nucleus which mainly cools down by nucleon emission, the latter completely disintegrates into a stream of light nuclear clusters.
	We define this mechanism as nuclear jet and, as discussed thereafter, we find that these clustered structures are not triggered by cohesive forces, which may only give a minor contribution, but they rather recall the granular flow of a stream of dry sand.
	Detailed experimental analyses of nuclear jets are very recent~\cite{Francalanza_inpreparation}, but signatures of this mechanism have been observed in many previous experiments without attracting the deserved attention, probably due to their difficult interpretation~\cite{Lautesse2006}.
	We argue in fact that the behaviour of a nuclear jet is a complex combination between volume instabilities and surface fluctuations as a function of the equation-of-state (EOS) coordinates.

	Such picture emerges when transposing the physics of nuclear matter to the phenomenology of HIC.
	In nuclear matter, along the EOS landscape, extreme values of density and incompressibility may result in unstable conditions, typically achieved below saturation density $\rho_{\textrm{sat}}$.
	An example is the spinodal instability which acts by amplifying a disturbance propagating in a volume.
	In these circumstances, a general property of the nuclear interaction is to induce clusterisation.
	In realistic conditions (HIC), however, matter organises into finite self-bound open systems which present surfaces and interfaces.
	Under the violent perturbations which characterise HIC, fluctuations propagate also along surfaces and take the form of a Plateau-Rayleigh instability~\cite{Brosa1990}, which provides a competing mechanism for clusterisation.
	Under the extremely violent perturbations which characterise nuclear jets, density drops to so small values that surface contributions are dramatically reduced and volume instabilities can prevail.

\section{Modelling volume and surface fluctuations: from nuclear matter to HIC}

	In order to analyse clusterisation as a process, as a function of density and time, allowing for non-equilibrium response to violent perturbations, we should search a microscopic description in terms of the underlying dynamics of fluctuations.
	This intricate mechanism can be studied through the use of one-body approaches, well suited to describe zero-sound conditions and collective effects, supplemented by higher-order correlations to introduce fluctuations.
	Fluctuations are then able to develop spontaneously and enhance the leading instability, either of volume or surface type, depending on incompressibility, density, temperature and symmetry energy.
	Well adapted to this purpose, the Boltzmann-Langevin-One-Body (BLOB) model~\cite{Napolitani2013} has been developed following the scheme of the above introduction, i.e. the requirement of describing dynamical instabilities in nuclear matter~\cite{Napolitani2017} and successively extending the approach to HIC~\cite{Napolitani2017_HDR,Colonna2017,Napolitani2016_NN2015}.

The BLOB approach is a technique explicitly suited for describing clusterisation as resulting from large-amplitude fluctuations acting in a semiclassical mean-field framework and intermittently revived by collisional correlations.
	In the spirit of a Brownian-motion~\cite{Chomaz1994} scheme, an ensemble of distribution functions $f^{(n)}$ are exploited, each one representing a possible mean-field trajectory, so that the BLOB approach may be seen as the Wigner transform of a stochastic TDHF approach extended to include a two-body collision term (see ref.~\cite{DeLaMota2019_IWM2018}).
	In particular, correlations beyond the kinetic-equation level (or, equivalently, the upper orders beyond two-body correlations in a BBGKY hierarchy~\cite{Lacroix2016}) are introduced in full phase-space continuously in time.
Such correlations are introduced by letting collide phase-space portions of extended size, large enough that the occupancy variance in $h^3$ cells corresponds to the scattering of two nucleons, and by simultaneously imposing that the  Pauli  blocking  of  initial  and final states is nowhere violated over the full phase-space extension.  
The BLOB approach solves therefore the Boltzmann-Langevin (BL) equation in full phase space.
For comparison, we also consider a second simplified `collisionless' approach to solve the BL equation in a drastically simplified form, where numerical noise is used as the only fluctuation seed.
In both approaches a  simplified SKM* effective interaction~\cite{Guarnera1996,Baran2005}, with a compressibility modulus $k=200$MeV, a linear form for the symmetry energy, and momentum-dependence omitted, is used.


	In nuclear matter (NM), in presence of unstable conditions, like negative incompressibility, we can associate to a point of the EOS a dispersion relation which links the form of the nuclear interaction to the growth rate of disturbancies of given wave number $k$ in the volume.
	The range of the nuclear interaction imposes an ultraviolet cutoff on the $k$ distribution, so that the dispersion relation is not displaying a linear growth of the fluctuation response amplitude as a function of $k$, but it goes to zero by excluding small wavelengths $\lambda$ as a function of the interaction range. 
	As a consequence, the dispersion relation presents a maximum (largest growth rate, or smallest growing time), so that the corresponding $k$ modes lead the instability.
	On the other hand, when regarding open and finite systems, surface contributions should be introduced, ruled by the same term of the nuclear interaction that was imposing the ultraviolet cutoff in the dispersion relation of nuclear matter.
	In this case, fluctuations act also on surfaces, imposing Rayleigh unstable behaviours in addition to the volume perturbation (i.e. spinodal instability and Landau damping).

	In heavy-ion collisions, to isolate volume perturbations, the system should be heated up in order to let the density $\rho$ drop to around a third of the saturation density $\rho_{\textrm{sat}}$~\cite{Borderie2018}.
	Experimentally, the most suitable configuration for studying volume fluctuations in HIC is an isotropically expanding collision remnant, achieved from maximum stopping in symmetric central collisions at Fermi energies~\cite{Chomaz2004,Colonna1997} (or in proton-induced spallation~\cite{Napolitani2015,Napolitani2016_IWM2016} and analogue mechanisms).
	On the other hand, suitable configurations where surface fluctuations dominate are neck-like structures formed in fissioning systems~\cite{Montoya1994,Toke1995,Lecolley1995}.
	Between these two extreme scenarios, HIC are therefore a battlefield where volume and surface instabilities compete with different contributions depending on the mechanism.

\section{Results: phenomenology of nuclear jets}

%
%
\begin{figure}[t!]\begin{center}
	\includegraphics[angle=0, width=.7\textwidth]{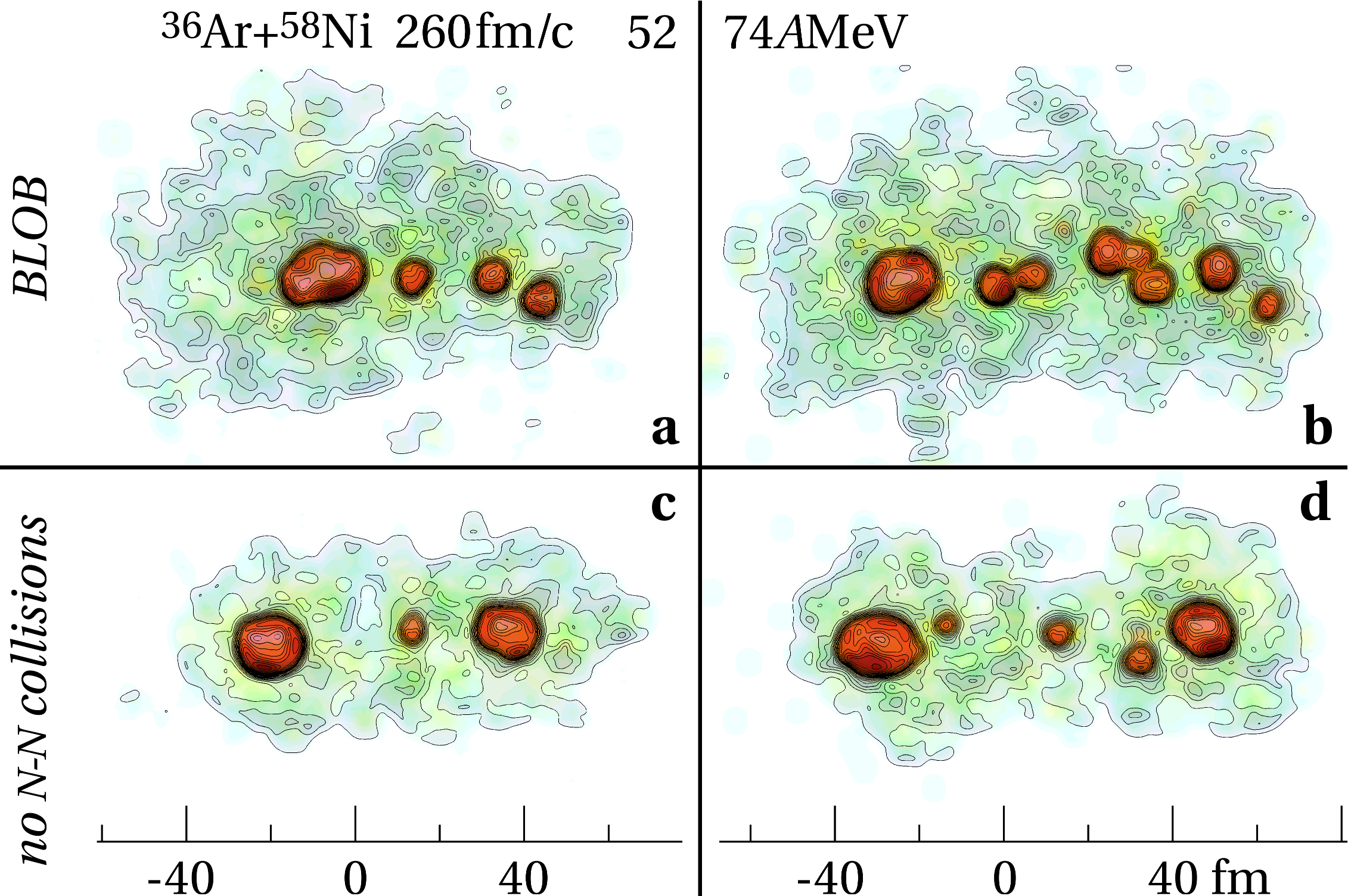}
\end{center}\caption
{
	Selection of most probable configurations at 260fm/c with many clusters 
	in $^{36}$Ar$+^{58}$Ni head-on collisions.
	{\bf a}, 52~$A$MeV, BLOB calculation: jet mechanism.
	{\bf b}, 74~$A$MeV, BLOB calculation: jet mechanism.
	{\bf c}, 52~$A$MeV, collisionless calculation: neck mechanism.
	{\bf d}, 74~$A$MeV, collisionless calculation: neck mechanism.
}
\label{fig1}
\end{figure}
	We briefly review some results about head-on $^{36}$Ar$+^{58}$Ni collisions at Fermi energies, taken as an example of system where nuclear jets were observed both experimentally~\cite{Francalanza_inpreparation,Lautesse2006} and in the corresponding simulations~\cite{Napolitani_submitted}.
	With increasing energy, the exit channel moves from mostly fusion (below 25~$A$MeV) to a large tendency towards the disintegration of the whole system into clusters and light particles (above 90~$A$MeV).
	In between, from 52 to 74~$A$MeV, in the BLOB calculation we observe the disappearing of the $^{36}$Ar projectile remnant and the production of a stream of clusters with element number ranging up to $Z\sim10$.
	Clusters are rather well aligned and they gradually turn into a more irregular flow (loss of alignment) for increasing energy. 
	This behaviour is illustrated in figs.~\ref{fig1}a,b, for the the systems $^{36}$Ar$+^{58}$Ni at 52 and 74~$A$MeV at 260~fm/c.
	In contract, figs.~\ref{fig1}c,d, show for the same systems that the corresponding collisionless approach would lead to a completely different exit channel: two large bulges which can be identified with the remnants of the participants nuclei $^{36}$Ar and $^{58}$Ni and one or more small neck-like fragments in between.

%
%
\begin{figure}[b!]\begin{center}
	\includegraphics[angle=0, width=1\textwidth]{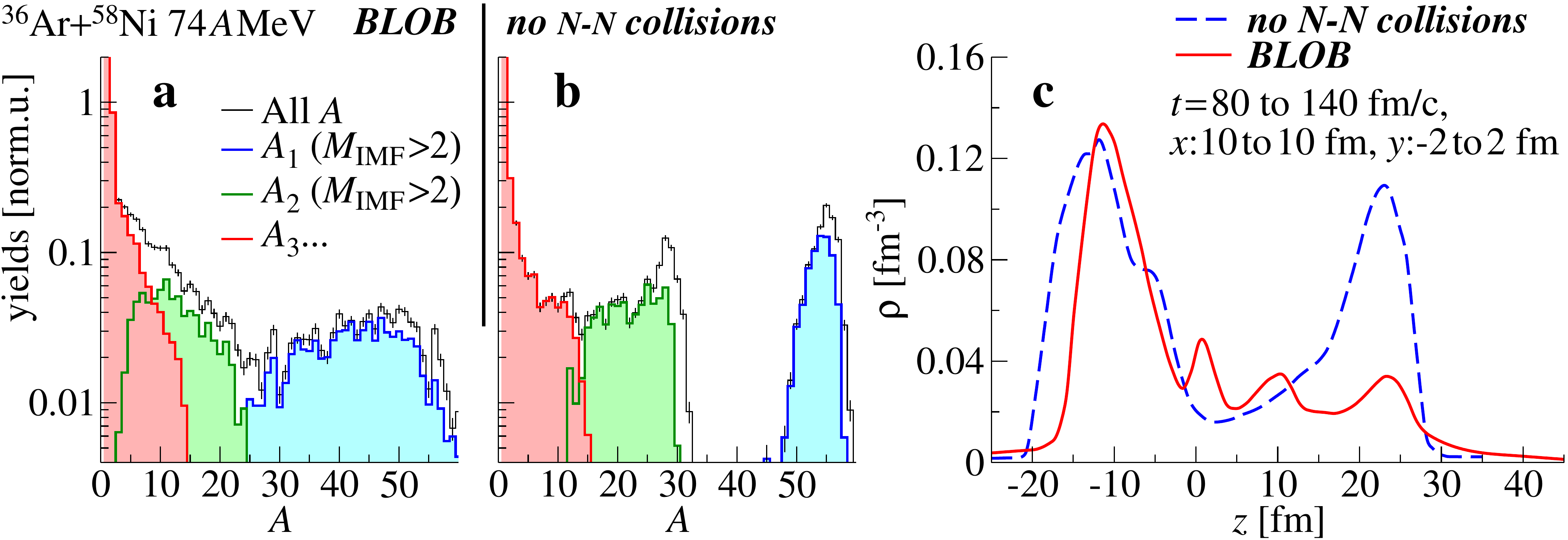}
\end{center}\caption
{
	Fragment production and average density survey
	in $^{36}$Ar$+^{58}$Ni head-on collisions at 74~$A$MeV.
	{\bf a,b}, Fragment production at the time of the last-fragment separation; BLOB ({\bf a}) and collisionless ({\bf b}) calculations.
	{\bf c}, Average density calculated in the time interval 80 to 140fm/c along the collision axis $z$ within the two model schemes.
}
\label{fig2}
\end{figure}
	Focusing on the system $^{36}$Ar$+^{58}$Ni at 74~$A$MeV, fig.~\ref{fig2} gives more insight about the mass distribution of clusters and fragments and their size hierarchy, through the separate analysis of the largest fragment $A_1$, the second largest $A_2$ and the remaining production $A_3...$
	In particular, fig.~\ref{fig2}a shows that in the BLOB calculation only two contributions to the distribution of reaction products can be distinguished, the target residue $A_1$ and the clusters composing the jet, without any possibility to further disentangle $A_2$ and $A_3...$
	On the contrary, fig.~\ref{fig2}b shows that in the collisionless approach $A_1$ and $A_2$ distributions are well distinguished and representative of the $^{58}$Ni and $^{36}$Ar residue production, respectively.
	Fig.~\ref{fig2}c, constructed for one single but most probable exit-channel adds information about the configuration space (longitudinal coordinate along the reaction axis $z$) and the density explored in the reaction right after separation of all fragments and clusters.
	The density distribution resulting from the collisionless approach exhibits two humps in forward and backward positions along $z$ connected by a low-density portion at midrapidity: the corresponding configuration is a low-density neck which forms at midrapidity between two heavy partners.
	In the BLOB calculation, on the other hand, the density distribution displays only one prominent hump, corresponding to the target remnant, and a long tail which drops to very low densities, even below $\rho_{\textrm{sat}}/4$, and which signs the developing of a nuclear jet.
	This latter result is a more reliable picture of the mechanism, not only because the calculation involves a complete treatment of fluctuations, but also because it is consistent with the experimental findings~\cite{Lautesse2006,Francalanza_inpreparation}. 

	Since the two approaches, BLOB and the collisionless counterpart, incorporate the same mean-field implementation, the dramatic difference manifested in the resulting mechanism should come from the different handling of fluctuations and dissipation contributions.
	To obtain a more direct insight on the situation, we should therefore analyse the fluctuation modes, in terms of wavelengths and growth times, distinguishing surface and volume contributions in the BLOB calculation.

	Schematically, we can approximate the forward part of the nuclear jet by a columnar configuration of radius $r$ at a given local density $\rho$, a temperature $T$, and an isospin $\beta=(\rho_n-\rho_p)/\rho$ for the emerging clusters, and obtain the growth rate $\Gamma_{k, \textrm{surf}}$ for surface fluctuations of wave number $k$ according to the following analytic prescription~\cite{Brosa1990}:
\begin{eqnarray}\label{eq:Rayleigh}
	(\Gamma_{k, \textrm{surf}})^2 &=& 
	\frac{\tilde{\gamma}(\rho,\beta,T)}{\rho m r^3} \frac{I_1(kr)}{I_0(kr)} kr (1-k^2r^2)
	\;,\\
	\tx{where\ }& &
	\tilde{\gamma}(\rho,\beta,T)
	\approx \alpha(T) \Big[
	1 -c_{\textrm{sym}}\beta^2 
		-\chi\Big(1-\frac{\rho}{\rho_{\textrm{sat}}}\Big)	\Big]
		\gamma_{\textrm{sat}} 
	\;.\notag
\end{eqnarray}
	$I_0$ and $I_1$ are modified Bessel functions and $m$ is the nucleon mass.
	$\gamma_{\textrm{sat}}$ is the surface tension at saturation density and zero temperature in symmetric matter, while $\tilde{\gamma}(\rho,\beta,T)$ is the corresponding quantity when taking into consideration the low density and isospin (using the prescription suggested in refs.~\cite{Iida2004,Horiuchi2017} for the SKM$^{*}$ interaction), and by further introducing the finite-temperature correction $\alpha(T)$ (using the prescription suggested in ref.~\cite{Ravenhall1983}, which at $T=3$~MeV yields a negligible correction $\alpha(T)\approx1$).

%
%
\begin{figure}[t!]\begin{center}
	\includegraphics[angle=0, width=1\textwidth]{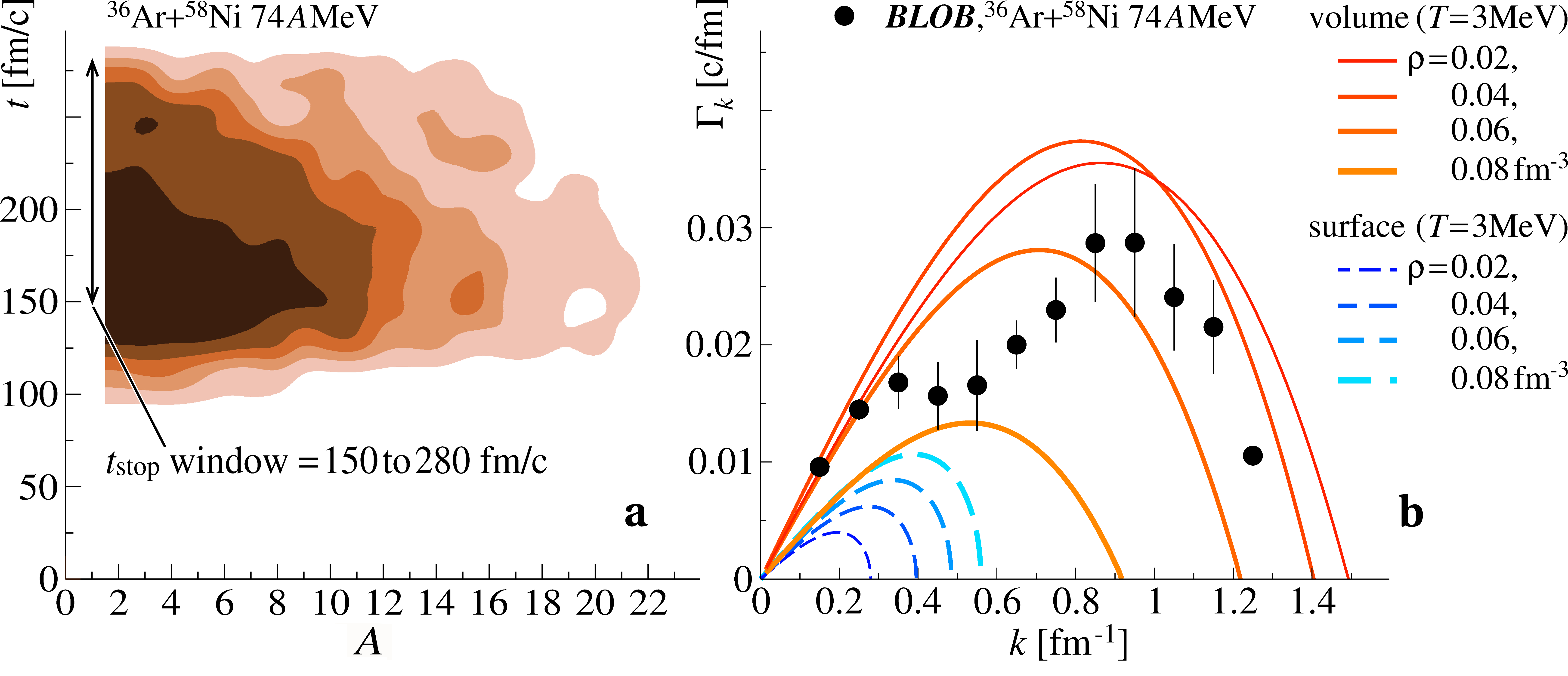}
\end{center}\caption
{
	Study of fluctuation contributions
	in $^{36}$Ar$+^{58}$Ni head-on collisions at 74~$A$MeV.
	{\bf a}, Breakup time of fragments and clusters as a function of their mass number, BLOB calculation.
	{\bf b}, Analytic spinodal (volume) and Plateau-Rayleigh (surface) dispersion relations calculated at the average temperature of the breaking-up system $T=$3 MeV for different densities; they are compared to the numerical estimation calculated with BLOB for the system $^{36}$Ar$+^{58}$Ni.
}
\label{fig3}
\end{figure}
	At the same time, we can also evaluate the growth rate $\Gamma_{k, \textrm{vol}}$ for volume fluctuations of wave number $k$ in conditions of spinodal instability within the linear-response approximation, as~\cite{Colonna1994_a,Colonna1994,Napolitani2017}:
\begin{equation}\label{eq:spinodal}
	1+\frac{1}{\tilde{F_0}(k,T)} =
		\frac{\Gamma_{k, \textrm{vol}}}{k v_{\textrm{F}}} 
		\textrm{arctan}\Bigg(
		\frac{k v_{\textrm{F}}}{\Gamma_{k, \textrm{vol}}} 
		\Bigg), \qquad
		\tx{where\ }
		\tilde{F_0}(k,T) = \frac{\mu(T)}{\epsilon_{\textrm{F}}} F_0 g(k)
	\;.
\end{equation}
	$v_{\textrm{F}}$ is the Fermi velocity and $\tilde{F_0}(k,T)$ is the effective Landau parameter including a dependence on temperature, through the chemical potential $\mu(T)$ and the Fermi energy $\epsilon_{\textrm{F}}$, and a modification $g(k)$ to take into account the nuclear interaction range.

	To the two alternative analytic expressions eq.~\ref{eq:Rayleigh} and eq.~\ref{eq:spinodal} we compare the dispersion-relation observables obtained directly from the BLOB simulation.
	They can be evaluated from a first-order analysis of chronology (as an analysis step, fig.~\ref{fig3}a illustrates the breakup time of fragments as a function of their size) and cluster correlations~\cite{Napolitani_submitted}, restricting to the forward section of the stream of clusters (i.e. the nuclear jet).
	The comparison of fig.~\ref{fig3}b, indicates that the fluctuation dynamics is not compatible with a pure scheme of surface instabilities of Plateau-Rayleigh type, but it is rather dominated by volume fluctuations, reflecting a very weak effect from surface tension in this mechanism.
	The surface effect seems however not to be completely negligible, as we may observe a reduction of the growth rate for large wavelengths (small $k$).
	The maximum of the dispersion relation is shifted to large $k$ modes (due to the contributions of different density values explored in the process), with the effect of favouring the production of light fragments and clusters.
	In particular, while the typical production from spinodal fragmentation should be situated in the region of Neon and Oxygen, the production from a nuclear jet should favour the region of deuterons, tritons, $^3$He and $\alpha$ particles.

\section{conclusions}

	Quite generally, nuclear processes should be regarded as the combination of two extreme behaviours, volume and surface fluctuations.
	In this spirit, we analysed the formation of nuclear jets and their disintegration into clusters.
	Such process, also observed in experiments, deliberately composes the conditions for volume instability (low density) and surface instability (large anisotropy).
	The BLOB approach describes consistently the jet mechanism because fluctuations and dissipation are implemented in full phase space.
	An approximate collisionless approach, where fluctuations are drastically reduced and simply projected on density space, and two-body correlation suppressed, would fail in describing such mechanism.
	We may mention that a similar study, leading to consistent results, is proposed also from comparing two versions of stochastic TDHF approaches~\cite{DeLaMota2019_IWM2018}.

	We found that the jet mechanism behaves favouring clusterisation from volume fluctuations, while it displays a weak dependence from surface tension.
Nevertheless, surface fluctuations, as well as the large range of densities involved, may affect the shape of the dispersion relation so that the largest growth rates tend to favour the production of clusters and fragments of small size.

\acknowledgments
Research was conducted in the scope of the International Associated Laboratory (LIA) COLL-AGAIN.

\end{document}